# Investigations on the Energy Balance in TDCB Tests

Olaf Hesebeck*, Udo Meyer, Andrea Sondag, Markus Brede

Fraunhofer Institute for Manufacturing Technology and Advanced Materials IFAM, Bremen, Wiener Str. 12, D-28359, Germany

* corresponding author, olaf.hesebeck@ifam.fraunhofer.de

## Abstract

The Tapered Double Cantilever Beam (TDCB) test is an established method to determine the critical strain energy release rate of adhesives in mode I. Provided that the adherends stay elastic, that the adhesive layer is not too flexible and that inertia effects can be neglected, the experiment allows to identify the work required by the adhesive layer per area of crack growth. The evaluation according to the standard does not permit to distinguish between different sources of dissipation in the adhesive layer or at the adhesive-adherend interfaces, though. This paper proposes two approaches to gain a more detailed understanding of the dissipation in mode I crack growth of adhesive layers.

The first investigation method uses detailed finite element simulations of the TDCB test based on an elastic-plastic adhesive material model derived from tests on bulk specimens. The simulation is used to distinguish between the work required for the plastic deformation of the entire adhesive layer and the work consumed by the crack and the adhesive in its vicinity. The dependence of this distribution of work on the adhesive layer thickness is studied. The second approach adds a temperature measurement by an infrared camera to the TDCB test. This measurement allows observation of the thermo-elastic effect in the adhesive layer and of the heat generation at the crack. Finally, the results of the two approaches are employed to estimate the energy balance in the TDCB test. The application to a ductile epoxy adhesive shows the feasibility of the proposed methods.

## Keywords

fracture toughness, finite element stress analysis, destructive testing, epoxides, heat generation

## 1 Introduction

Fracture mechanical tests are an established tool to characterize adhesives and to provide parameters for numerical modelling of fracture processes in adhesive joints [1]. The tapered double cantilever beam (TDCB) test is a standard test to determine the critical strain energy release rate for mode I crack growth of adhesives [2],[3],[4]. The specimen is designed so that the specimen compliance is a linear function of the crack length, if the effect of the adhesive layer compliance can be neglected. Consequently, the crack grows at a constant force during the test provided that the crack growth is stable and occurs at a constant energy release rate. This critical strain energy release rate $G_{Ic}$ can be evaluated from the square of the force according to the Irwin-Kies equation.

The critical energy release rate evaluated from the test is the ratio of the external work performed on the specimen minus the elastic energy of the adherends to the area of crack growth. This work may contain the work required to create the crack itself as well as work for inelastic deformation and damage in the entire adhesive layer. The evaluation of force-displacement curves of the test allows not to distinguish between the work required for different processes, but just provides the total $G_{Ic}$.

The contributions of plastic dissipation and the “intrinsic” work of fracture to the total work were already analyzed in the 1990s by numerical simulation. Those early works did not specifically regard adhesive joints. Tvergaard and Hutchinson [5] modelled crack growth at an interface between an elastic and an elastic-plastic solid using a cohesive zone model (CZM) at the interface. Assuming a mode independent CZM they showed that a mode dependence of the fracture toughness is caused by different amounts of plastic deformation outside the fracture process zone. A related approach assumed a small plasticity-free region close to the crack [6],[7].

An application of the approach to an interfacial crack in an adhesive joint between semi-circular, elastic adherends was presented by Chowdury and Narasimhan [8]. The Drucker-Prager model for the plasticity of the adhesive layer was combined with a trapezoidal traction-separation law at the interface. Madhusudhana and Narisham [9] simulated compact tension shear tests with a crack in the center of the adhesive layer for different loading angles and adhesive layer thickness.

A distinction between the intrinsic work of fracture on one side and the plastic dissipation and stored elastic energy in the adhesive layer on the other side was the aim of a work of Pardoen et al. [10] simulating wedge-peel tests. They combined a trapezoidal traction-separation law for the fracture process zone with von Mises plasticity in the adhesive layer. Using this approach, Martiny et al. [11] were able to model different configurations of the wedge-peel test of the adhesive Dow Betamate 73455 using a single set of model parameters. In particular, the dependence of the fracture toughness on the adhesive layer thickness (between 0.1 and 1 mm) could be explained by the contributions of adhesive plasticity and locked-in elastic energy while the fracture energy of the CZM was considered constant.

Next, Martiny et al. applied the same approach to TDCB tests of the adhesive Bondmaster ESP 110 which is tougher than the Betamate 73455 [12]. In this case it was not possible to simulate the tests of different adhesive layer thickness using a constant set of CZM parameters. Because of the difficulty to determine CZM parameters depending on stress triaxiality uniquely, Martiny et al. suggested to use a crack instead of the CZM. They employed a critical stress at a distance criterion to govern the crack growth.

After the submission of the current paper, Jokinen et al. published simulations of double cantilever beam (DCB) tests using an elastic – ideally plastic material model to describe the adhesive layer and the virtual crack closure technique (VCCT) to model the propagation of a crack [13]. They identified the critical energy release rate of the crack growth law iteratively by fitting the simulated force-displacement curve experimental data. The method exhibited computational challenges, and a dependence of the results on stabilization, fracture tolerance and time increment was observed.

The current paper suggests two approaches to gain more detailed information about different contributions to the energy balance in the TDCB test. Since the constant force in the TDCB test implies that no further information than $G_{Ic}$ can be obtained from the force-displacement curve, additional measurements are necessary to achieve this aim. The first proposed method follows a computational approach similar to the publications mentioned in the preceding paragraphs. It uses experimental data from tests on adhesive bulk specimens to create a material model. This model is used in finite element simulations which consider crack growth as well as the non-linear material behavior of the adhesive. The results are analyzed to learn how much the dissipation at the crack and in its vicinity and how much the inelastic deformation in the entire adhesive layer contribute to the critical energy release rate, respectively. The modelling method is similar to [12] and [13] regarding the description of fracture by crack growth instead of using a CZM, but we use a prescribed crack-growth velocity instead of a critical strain energy release rate or of a stress at a distance criterion. Furthermore, the rate-dependence of adhesive plasticity is considered.

The second proposed method increases the data gained from the TDCB test itself. A high speed infrared camera is used to observe the temperature change in the adhesive layer during crack growth. The resolution allows to distinguish between a temperature change in the entire layer, which can mainly be attributed to the thermo-elastic effect, and a local heat generation at the crack.

The results of the simulations and the thermal measurements are combined to make an estimate for the energy balance in the TDCB test. The most important underlying assumptions and open questions concerning their validity are pointed out.

The investigations have been performed using the construction steel S235 for the adherends and a cold-curing to part epoxy adhesive, Dow Betamate 2098. A TDCB specimen geometry according to Figure 1 with a specimen width of 20 mm and an initial crack length of 80 mm was used.

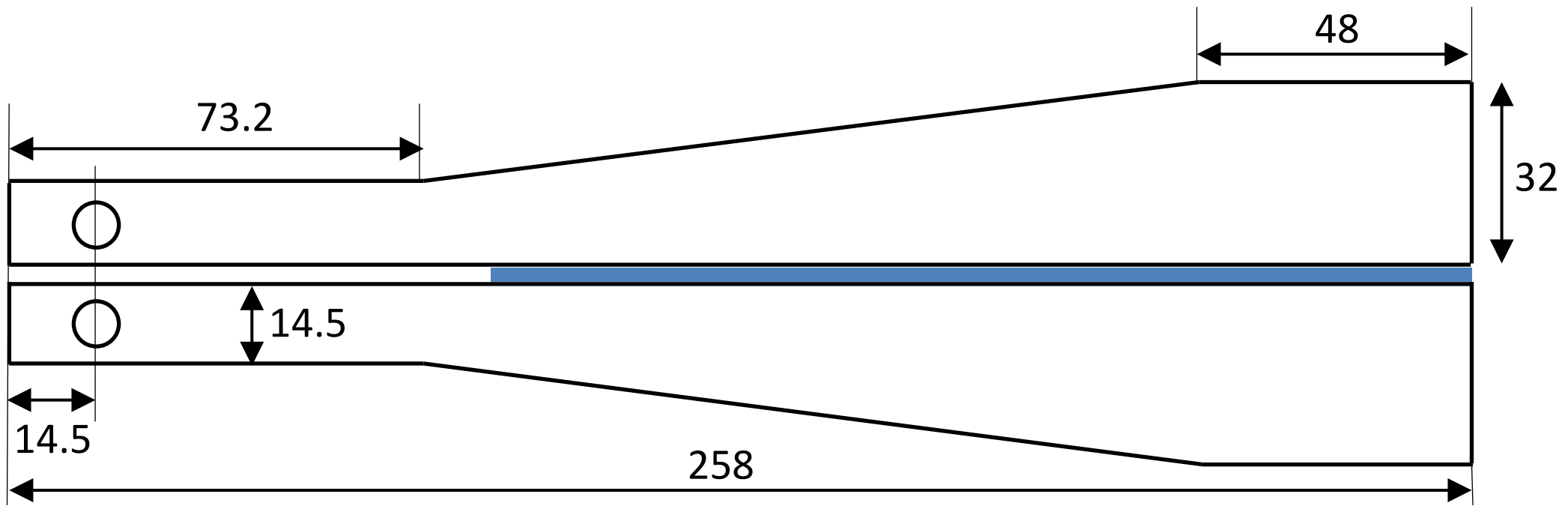

**Figure 1: Geometry of TDCB specimen**

# 2 Estimate of plastic work using finite element simulation

## 2.1 Tensile tests and material model

In this section additional information from tensile and compressive tests on the adhesive is used to learn about the energy balance in TDCB tests by finite element analysis. Tensile specimens according to DIN EN ISO 527-2 have been manufactured from the adhesive Betamate 2098. The specimens were tested at constant strain rates of 0.005, 0.05 and 0.5 1/s. These rates were chosen according to the strain rates encountered in the simulations of the TDCB tests. Additionally, tensile tests with unloadings were performed (Figure 2). A complex material behavior is observed: The hysteresis loops as well as the rate dependence of elastic stiffness indicate a visoelastic behavior. The decrease of slope of the hysteresis loops can be described by a damage mechanics model. The strain remaining after the unloadings can be modelled phenomenological using an elastic-plastic model. In the simulation of the TDCB test, the adhesive is loaded as the crack advances closer to the considered material point and unloaded after the crack has passed. For this kind of loading and the aim to estimate the work performed in the deformation of the adhesive layer, an elastic-plastic material model is the best choice, although it is not capable of capturing all properties of the adhesive. The hardening function is defined rate dependent based on the tensile tests performed at different strain rates. A simple von Mises yield function is chosen. While epoxy adhesives often require a yield function depending on the hydrostatic stress, the Betamate 2098 showed only a minor difference (8%) between the yield stress in tension and compression.

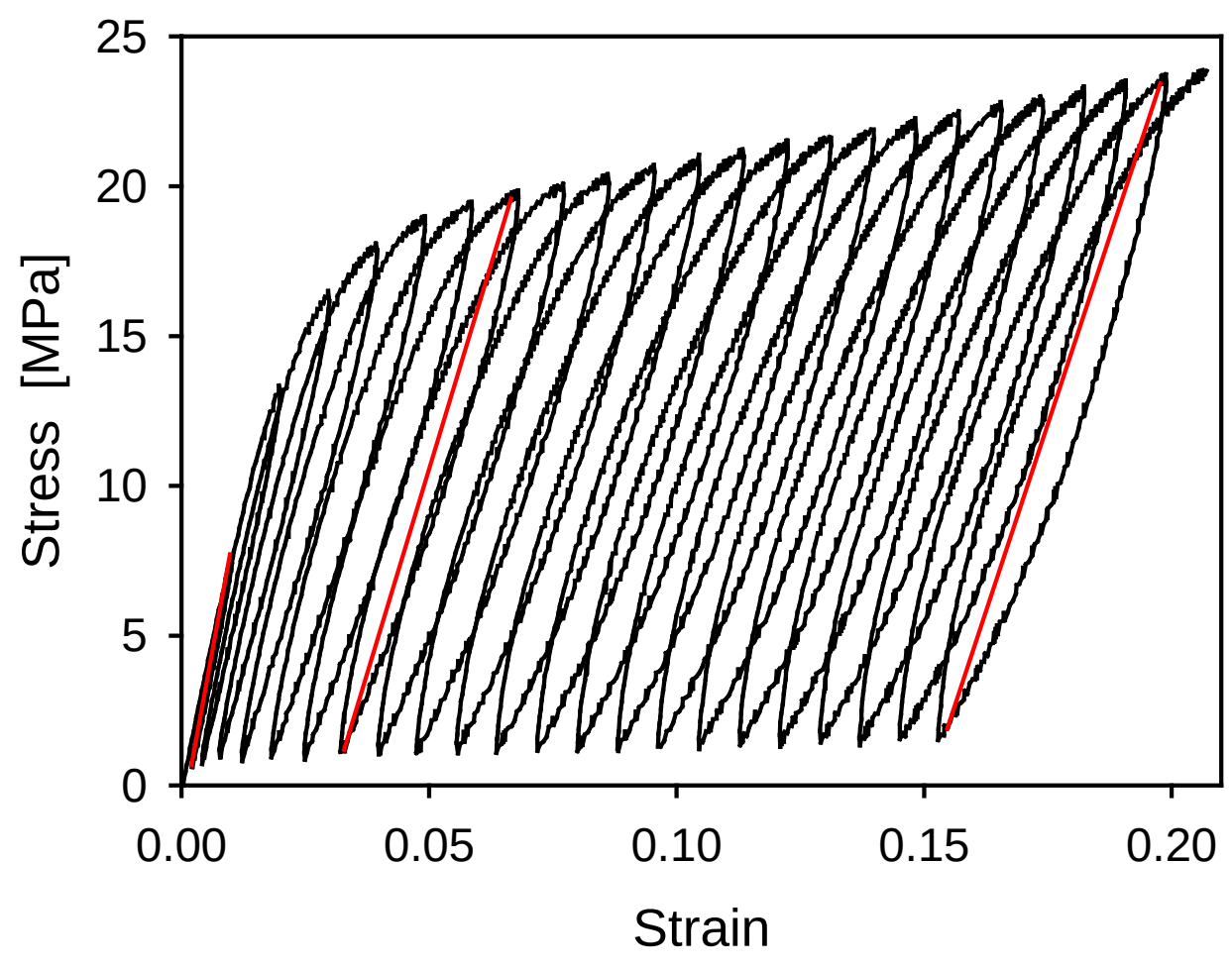

Figure 2: Stress-strain curve of tensile test with unloadings, Betamate 2098

The adherends made from steel S235 are considered as linear elastic.

## 2.2 Simulation and evaluation of plastic work

After identifying the adhesive material model, a finite element model of the TDCB test is set up. A crack is placed either inside the adhesive layer or at the interface to one adherend. In either case the crack extension direction is prescribed being parallel to the adhesive-adherend interfaces, which is a significant simplification of reality. Two different possible representations of the crack have been considered: a cohesive zone model and a fracture mechanics model.

The cohesive zone model uses a damage mechanics description for a thin layer representing the crack and its vicinity within the adhesive layer. Compared to the fracture mechanics approach it offers a more realistic model of ductile adhesives by accounting for a softening zone ahead of the crack. Its disadvantage is the need to identify at least three model parameters which are not directly accessible from the TDCB test results: stiffness, strength and critical energy release rate of the cohesive layer. Maybe the elastic modulus measured in the tensile tests can be used to calculate the stiffness, but the thickness of the cohesive layer still remains an arbitrary choice. It seems unlikely if not impossible to identify the required parameters uniquely, unless we assume that the same parameters hold for any adhesive layer thickness and use test results for several values of thickness for identification. Since the work of Martiny et al. [12] indicates that parameters independent of the adhesive thickness cannot be found for all adhesives, the cohesive zone approach was tested but finally not selected for the current investigation.

The fracture mechanical view considering a sharp crack in virgin material is more simplifying than the cohesive zone model, but offers the distinct advantage of requiring less parameter to be identified. A straightforward implementation in finite elements can be performed using the virtual crack closure technique (VCCT) with a critical energy release rate governing the crack growth. It should be noted that the critical strain energy release rate is lower than the value identified by the TDCB test, because the latter contains additionally the work spent for plastic deformation of the adhesive layer. Therefore, one parameter remains that needs inverse identification from the force-displacement curves measured in the TDCB tests. Since the VCCT models show a poor convergence behavior and a fine mesh is required to represent the deformation in the adhesive layer properly, this inverse identification has a high computational cost.

Therefore, an alternative approach for the fracture mechanics modelling of the crack in the TDCB specimen was developed here. While the critical energy release rate criterion is capable of predicting crack growth, this capability is actually not required for the current task. The TDCB tests shall not be predicted but evaluated using the simulation. Consequently, the most simple and efficient approach is to control the crack growth in the model by prescribing the crack tip position as a function of time. The crack growth in the model has to be chosen so that the simulated force during crack growth agrees with the experiment. It should be noted that the crack length in the simulation will not exactly be equal to the length of the real crack, but has to be considered as an effective crack length, because the damage in the vicinity of the crack tip is not considered by the fracture mechanics model.

If the response of the TDCB specimen is linear, the actual crack length can be calculated from the force and displacement. The TDCB specimen geometry is designed so that the specimen compliance *C* is a linear function of crack length *a*. If we determine the compliance by linear FEA for two different crack lengths, we have identified the linear relation between compliance and crack length. In particular, it can be shown that the crack growth velocity is a simple function of loading velocity *v* and force *F*:

$$\frac{da}{dt} = \frac{v}{F\frac{dC}{da}} \tag{1}$$

In this way it is possible to define the crack growth as a function of time in a way that the force at crack growth in the simulation of the test fits the force measured in the experiment. While the assumption of linear specimen behavior is fulfilled well for brittle adhesives or thin adhesive layers, for the adhesive Betamate 2098 and 0.6 mm adhesive layer thickness it leads to a deviation of 2% between the simulated force during crack growth and the desired value. This deviation is small compared to the experimental scatter, but detrimental to numerical studies like the investigation of the effect of crack position or element size. Therefore, the time control of crack growth was slightly shifted based on a non-linear FEA without crack growth.

The TDCB specimen was modelled in two dimensions using plain strain conditions. Figure 3 shows an example of the employed finite element meshes. The crack is located at the interface between adhesive and adherend in this example. Crack positions in the center of the adhesive layer and near to the interface have been considered as well. The mesh is locally refined to observe the adhesive layer in detail during a short period of crack growth. The smallest element edge length used to study the mesh influence was 6 µm, so that the model contained 830367 degrees of freedom. Typically, a few thousands time increments were needed to simulate the crack growth through the region of interest. The finite element software Abaqus was used.

The load was applied by giving the displacement of the holes in the adherends as a function of time while allowing free rotation.

The equivalent plastic strain was evaluated in the simulations. Since the investigated adhesive is highly non-linear, the plastic zone advances about 50 mm ahead of the crack tip for 0.6 mm adhesive layer thickness. A short distance behind the crack tip the plastic deformation is completed and the equivalent plastic strain does not change any more, see Figure 4. To calculate the work spent for plastic deformation the plastic strain is evaluated along a section crossing the adhesive layer in the direction of adhesive thickness. This is displayed in Figure 5 for different meshes denoted by their respective element size in x- and y-direction.

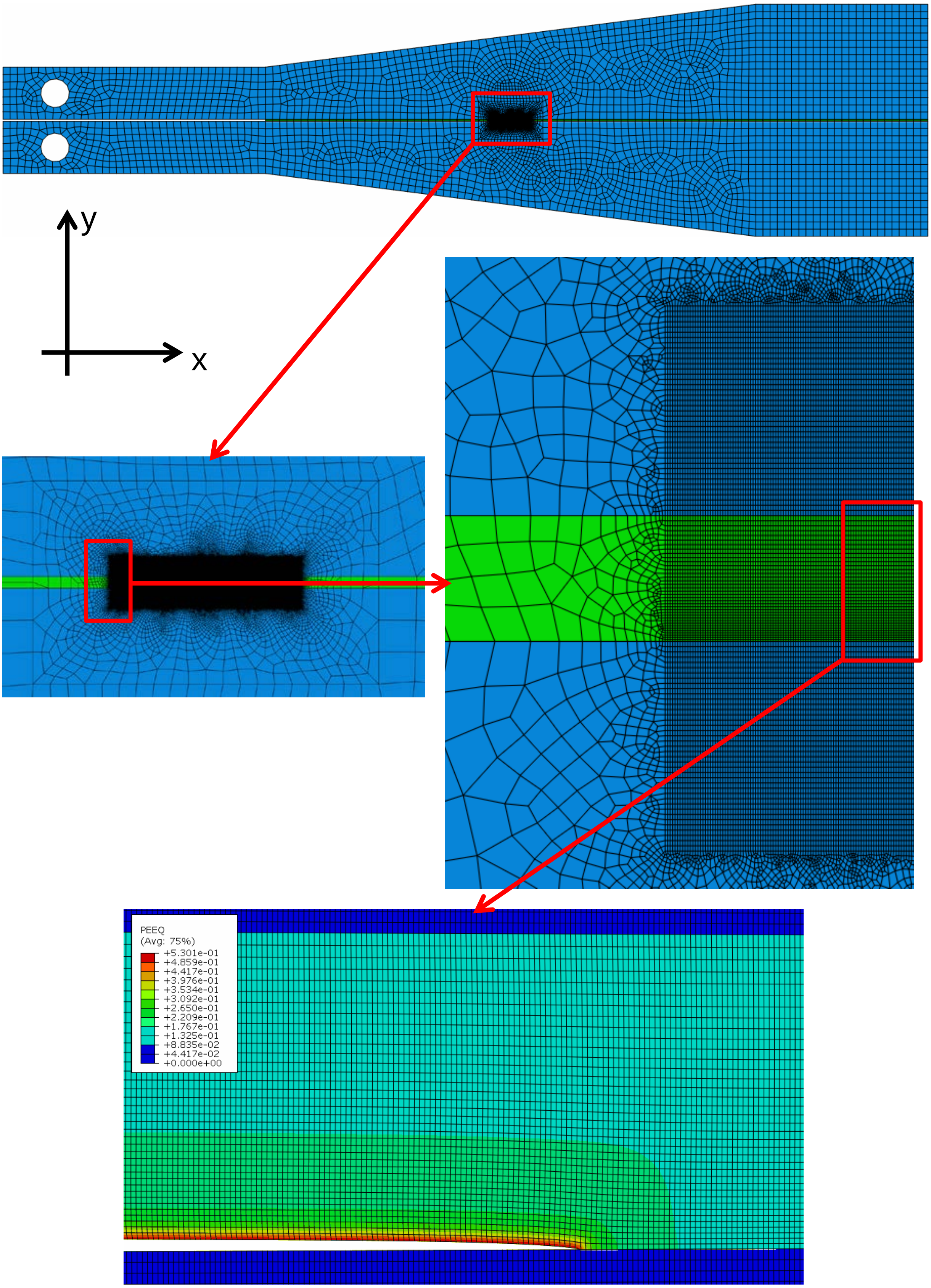

**Figure 3: Example of finite element mesh, adhesive thickness 0.6 mm**

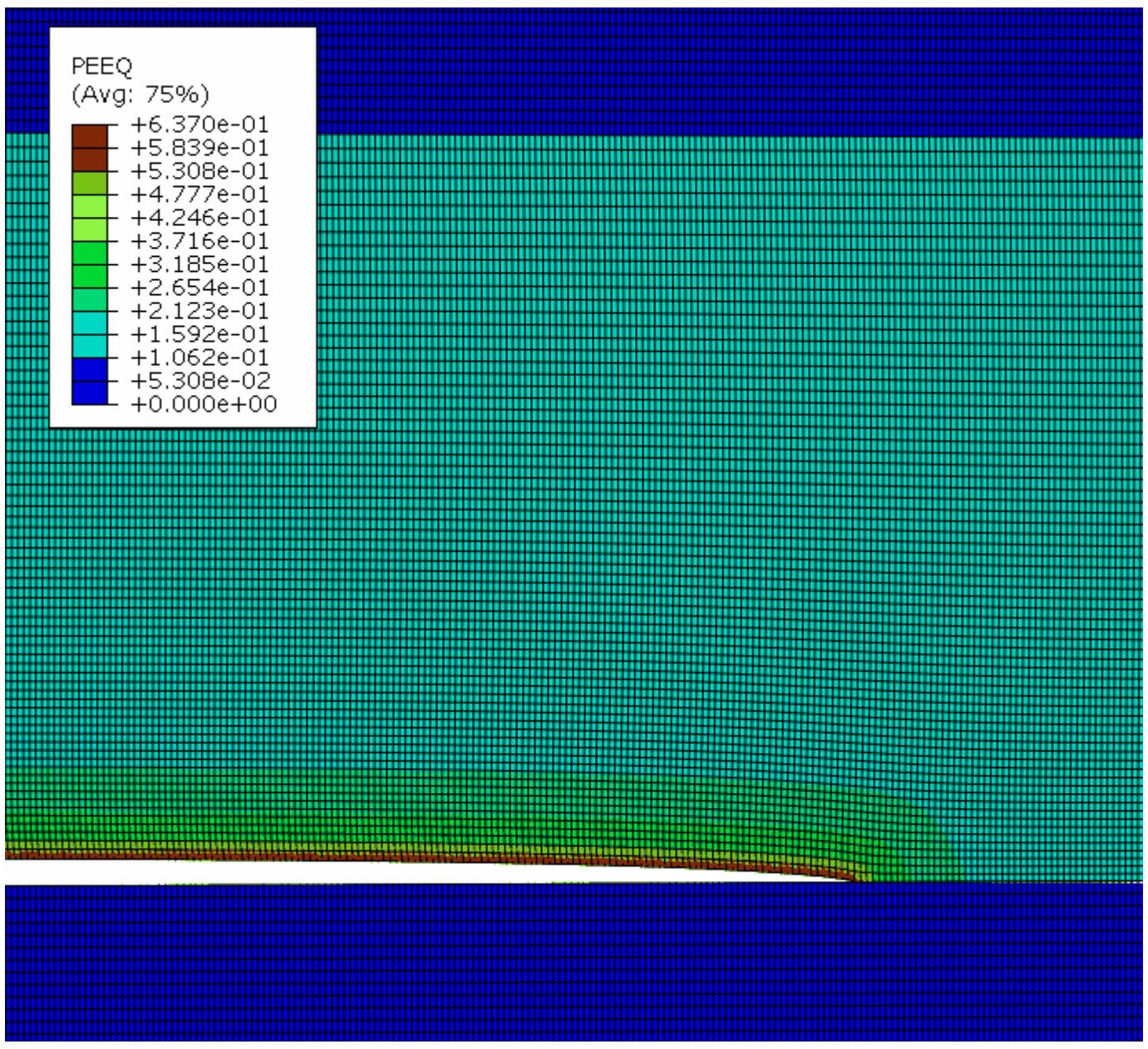

**Figure 4: Equivalent plastic strain near crack tip**

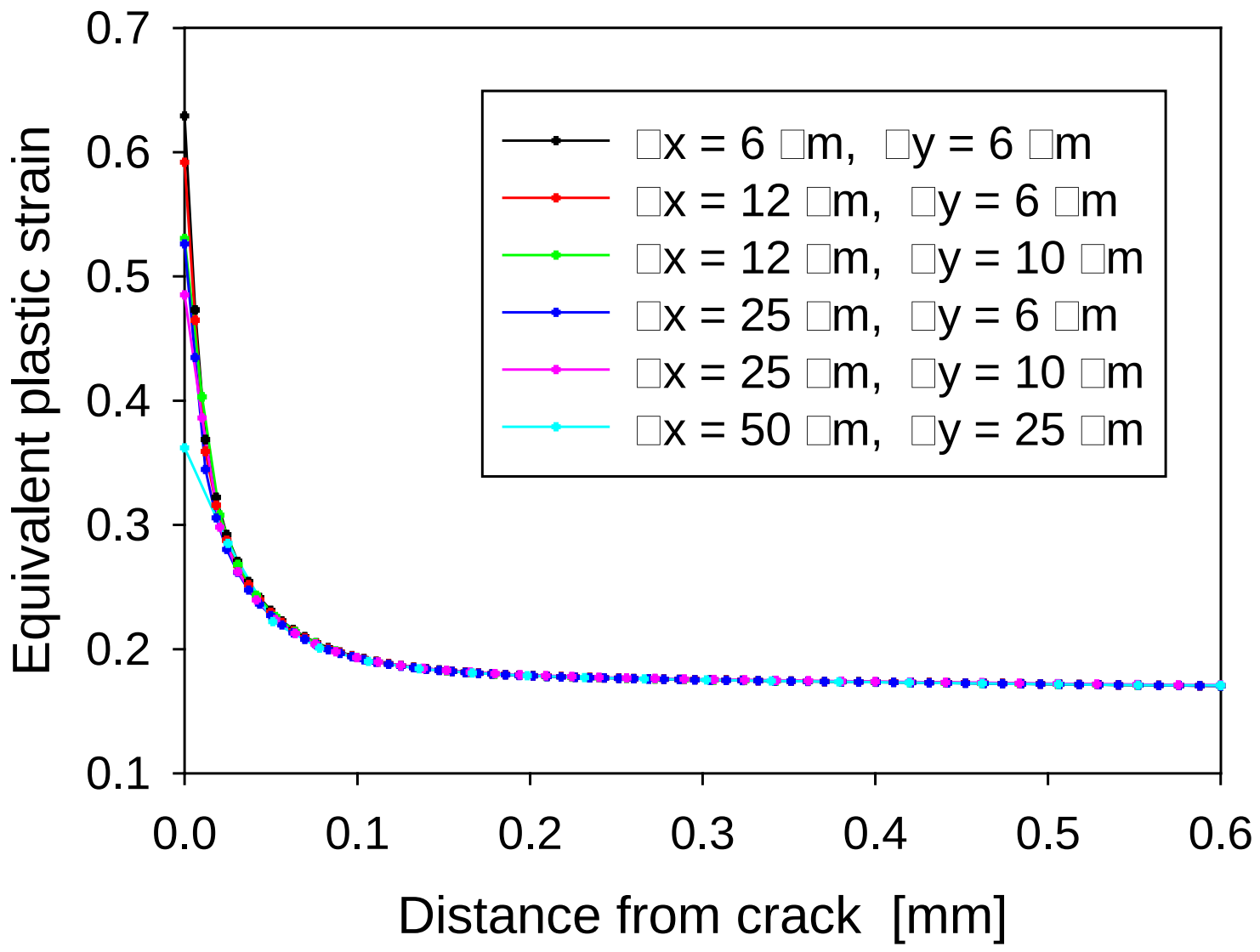

**Figure 5: Equivalent plastic strain as a function of distance from the crack**

The results are fairly mesh independent apart from the first elements close to the crack where the highest plastic strains occur. The plastic strain near the crack of about 0.6 exceeds the maximum strain observed in the tensile tests used to identify the adhesive material model. Figure 6 show the stress-strain curves of tensile tests performed at a strain rate of 0.05 1/s. the largest plastic strain reached in the tests was 0.3. Beyond this value, the material model uses a linear extrapolation of the hardening. While this looks like a reasonable extrapolation, it is certainly not valid for arbitrarily large deformations. Furthermore, the results of the simulations are questionable if they depend significantly on the details of the extrapolation. Therefore, a modified material model with less hardening in the extrapolation range (dashed line) was employed for comparison.

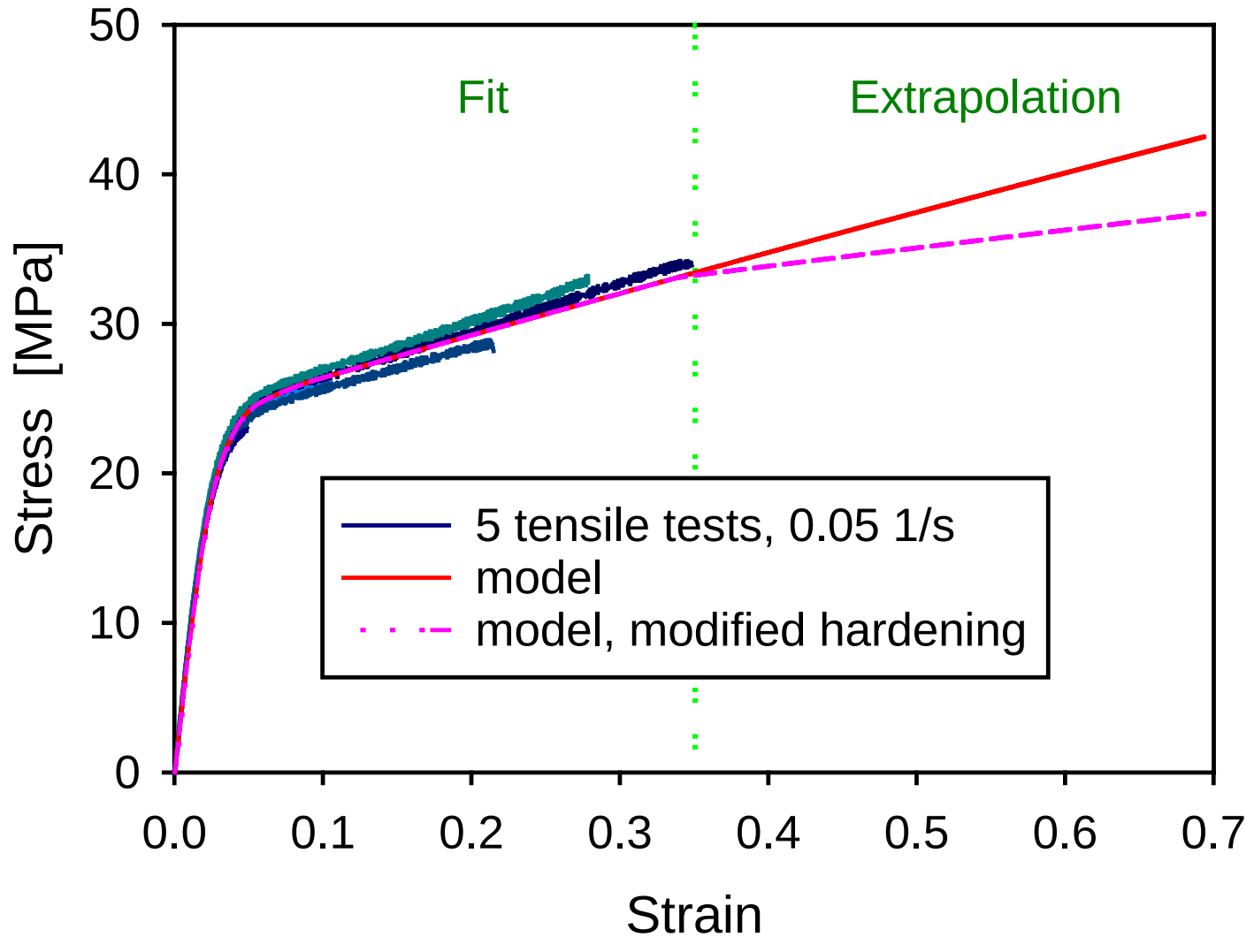

**Figure 6: Tensile tests and model extrapolation**

The result of this comparison is shown in Figure 7. For distances to the crack larger than 25 µm the strain stays below 0.3 and the simulations show no significant differences. Closer to the crack where the plastic strain exceeds the maximum value of the tensile tests, the different material models lead to different results as expected. This means that despite of our lack of knowledge about the adhesive behavior at high strains, we still get reliable results in the part of the adhesive layer where the plastic strain stays below the limit of 0.3. Therefore, we are able to calculate the work spent on plastic deformation of that region which is shaded in green in Figure 8. The difference between the total $G_{Ic}$ as obtained from the classical evaluation of the test and this value of plastic work in the green shaded part of the adhesive bulk yields the work used at the crack and its vicinity.

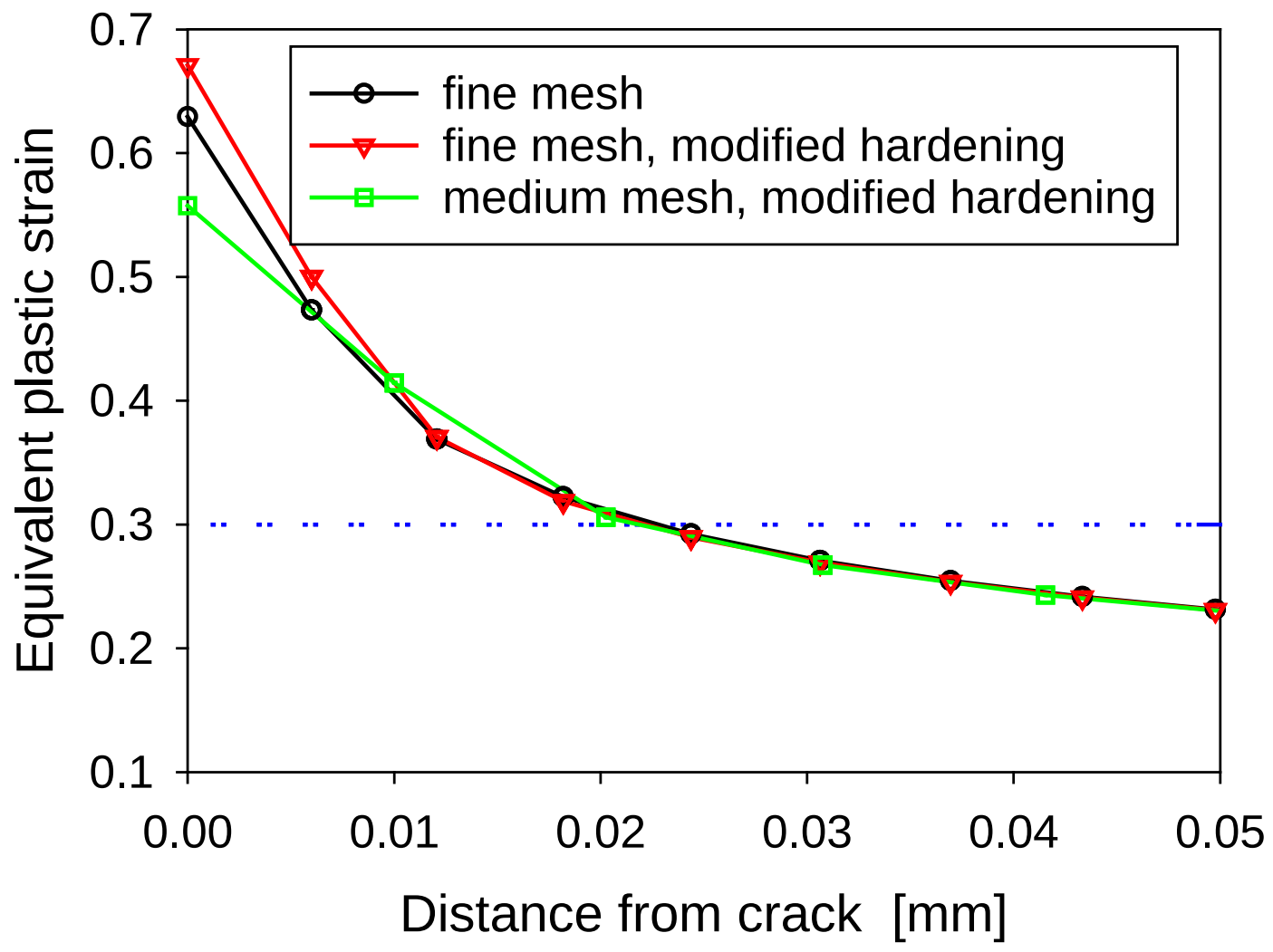

**Figure 7: Effect of material model on simulation results**

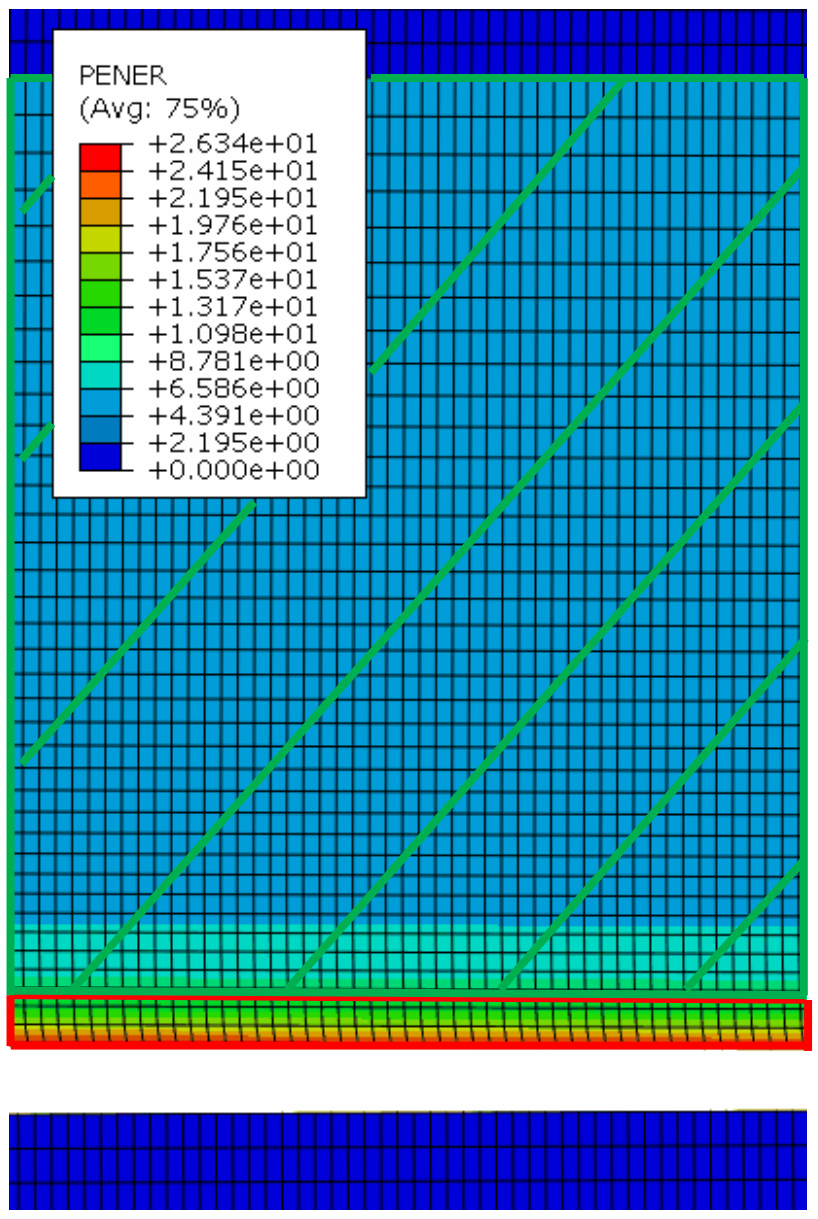

Figure 8: Vicinity of crack (red) and major part of adhesive bulk (shaded green), displayed on contour plot of plastic energy density

This kind of evaluation was performed for different meshes and using the original modified material model. The results are displayed in Figure 9. The circles connected by the green line show the ratio of the plastic work spent in the green shaded area to the total $G_{Ic}$ which shall be called "plastic fraction" for abbreviation. This value is about 0.8 quite independent of the mesh refinement and the material model in the extrapolation range. The triangles connected by the red line display the thickness of the zone at the crack exceeding 0.3 equivalent plastic strain. This quantity shows some mesh dependence which is to no surprise considering its small size about 20 µm.

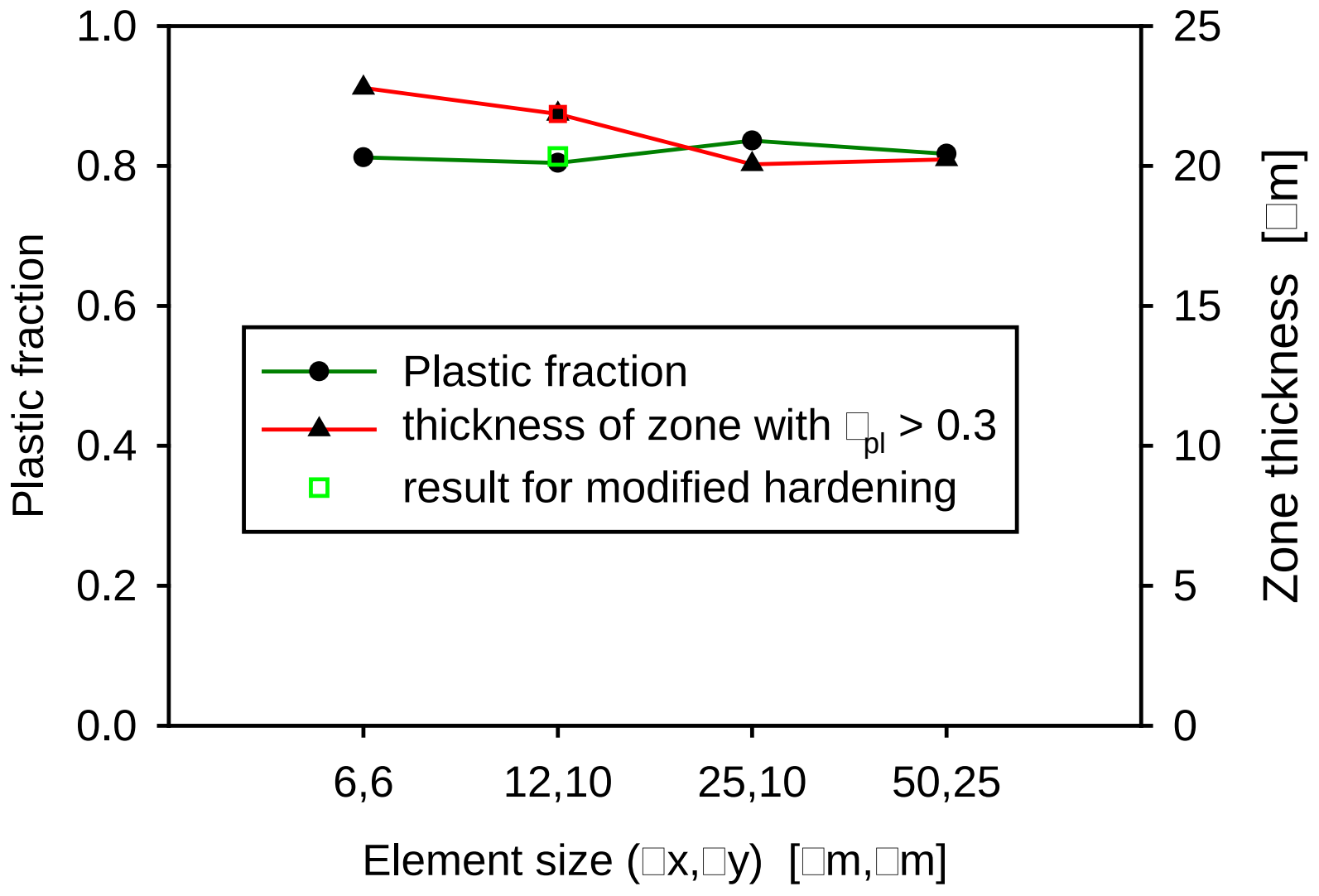

Figure 9: Mesh and material law dependence of results

## 2.3 Evaluation of TDCB test results using the simulation

TDCB tests of the adhesive Betamate 2098 were performed for different thicknesses of the adhesive layer. The tests velocity was 1 mm/s for an adhesive thickness of 0.3 mm and scaled proportional to the thickness for adhesive thicknesses of 0.6, 1 and 2 mm. 5 specimens per thickness were tested. The measured force-displacement curves in Figure 10 show stable crack growth.

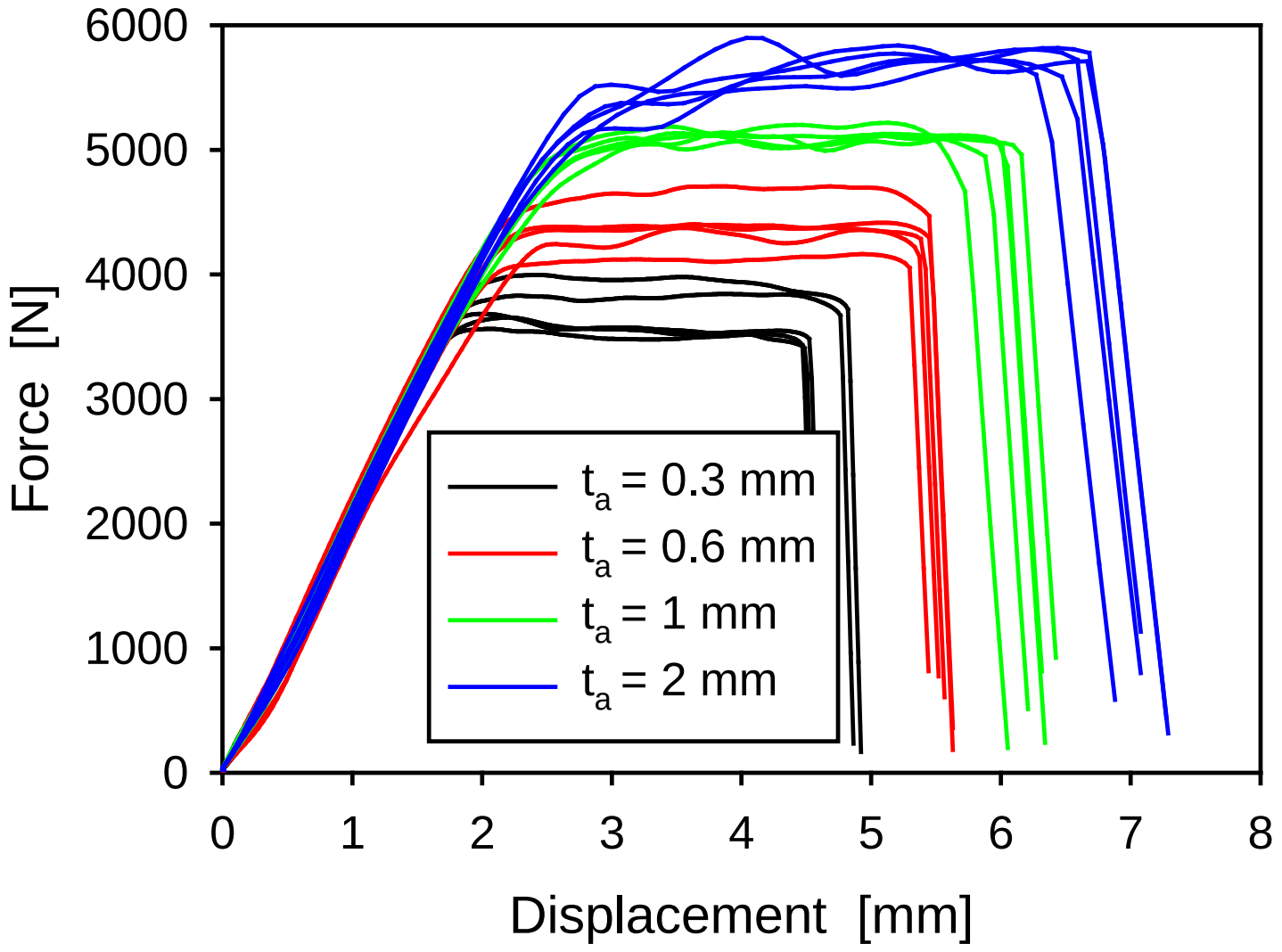

**Figure 10: Force-displacement curves of TDCB tests at different adhesive thickness**

For each adhesive layer thickness a simulation according to the previous section was set up so that the simulated force during crack growth is equal to the mean of the values observed in the tests. The total work excluding the elastic energy of the adherends per area of crack growth and the contribution of plasticity were evaluated from the simulations. This total work agreed with the value of $G_{Ic}$ obtained directly from the tests using the experimental compliance method with accuracy about 6%.

Figure 11 shows an increase of $G_{Ic}$ obtained from the simulations with increasing thickness (black line) which is the expected behavior for this kind of adhesive. The simulations enable us to split this total work required for the crack growth into the energy dissipated at the crack and its vicinity (red curve) and the plastic work performed in the major part of the adhesive layer (green curve). Both parts increase with increasing thickness. However, the plastic fraction (blue diamonds) stays approximately constant around 0.8.

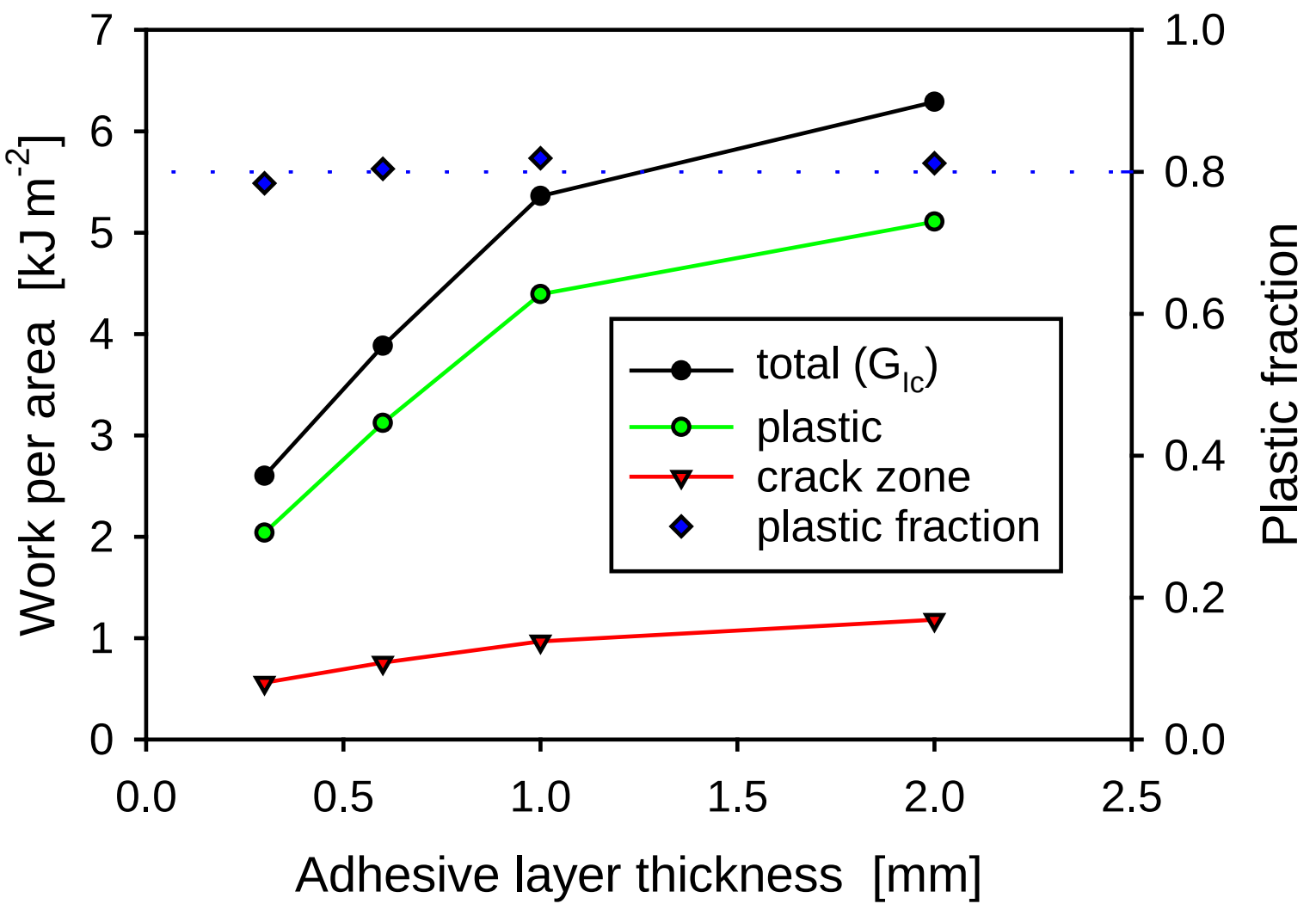

Figure 11: Dependence on adhesive layer thickness

While these first results obtained by the new analysis method indicate that the plastic fraction could be a property independent of adhesive layer thickness, the value certainly depends on the nature of the adhesive. A similar simulation and evaluation was applied to a much more brittle adhesive (Hilti Hit RE 500) of 0.6 mm thickness and calculated a plastic fraction of about 0.1.

# 3 TDCB-tests with infrared camera

While the simulations shown in the previous section basically used the additional information from tensile tests to learn more about the energy balance during crack growth in the TDCB test, an additional measurement during the TDCB test will be considered now. A high speed infrared camera (Taurus 110 K) was used to observe the temperature of the free surface of the adhesive layer on one side of the specimen. The aim of this investigation was to learn how much of the work in the test is dissipated in heat and if this heat generation is localized at the crack or occurs in the entire adhesive layer.

## 3.1 Experimental observations

Four of the pictures taken by the infrared camera during a test performed with Betamate 2098 and 3.6 mm adhesive layer thickness are displayed in Figure 12. The test was performed at a relatively high test speed of 8 mm/s, so that it was possible to observe temperature changes before the heat conduction evens out the temperature gradients. The color range from black, blue, red, and yellow to white indicates increasing thermal emission. The adherends appear mostly black due to the low thermal emission coefficient of the steel. The first picture taken at the time 3199 ms shows a state when the part of the adhesive layer observed by the camera is still nearly unloaded. At 3380 ms we see a rigid body movement of the specimen. Closer inspection reveals that the adhesive layer is strained in thickness direction at this stage. The blue color indicates that the temperature in the adhesive has decreased. In the next picture, a crack has appeared from the left side. It continues to grow while the temperature in the main part of the adhesive layer rises again close to its initial value. At 3684 ms just before the total fracture of the specimen the crack has opened considerably, and a locally increased temperature close to the crack is visible.

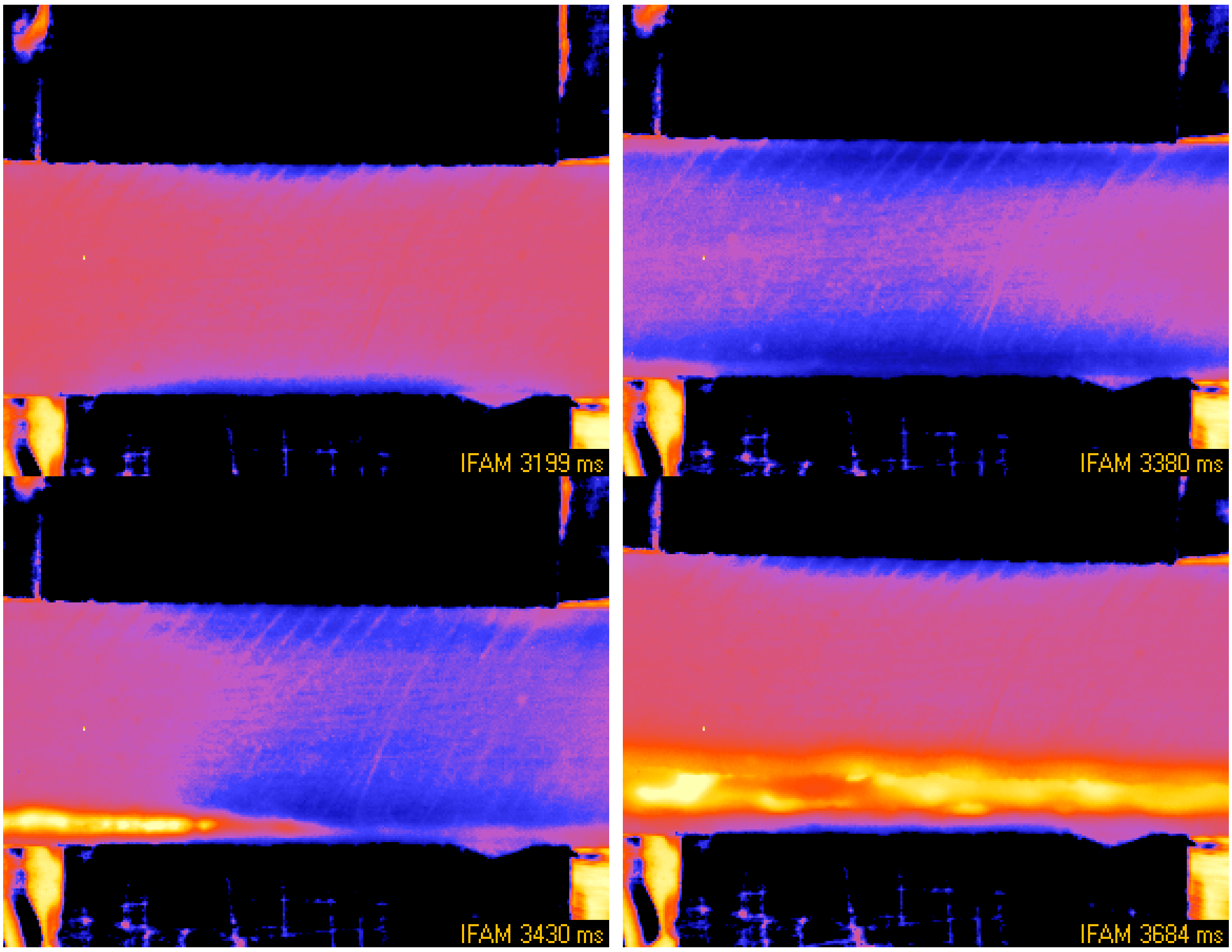

**Figure 12: Pictures from infrared camera during the TDCB test**

For a quantitative evaluation of the measurements the temperature profiles along several sections in thickness direction were extracted at several time points. Figure 13 shows these profiles for one section. In the center there is the adhesive layer with a temperature about 30 °C. On the left and right side the temperature appears lower due to the different emission coefficient of the adherends. A temperature peak in the adhesive layer is visible in the curves obtained from the images after the crack has reached the section.

Three quantities evaluated from the temperature profiles are displayed as a function of time in Figure 14. The blue line shows the current distance between the adherends. First it remains approximately constant. Starting at 3300 ms the adhesive layer is strained. The strain rate increases much when the crack tip reaches the section. The temperature in the center of the adhesive layer is displayed by the red line. Since the adhesive layer is moving relative to the camera, it is impossible to follow one material point on the adhesive surface exactly. Instead, the temperature was evaluated at a fixed distance from the adherend which is less close to the crack. This value is supposed to be a good approximation for the temperature experienced by a fixed material point. Its evolution is typical for the entire adhesive layer excepting the temperature peak at the crack. With increasing strain of the adhesive layer, this temperature decreases. After the crack has passed the section, the adhesive layer is unloaded again and the temperature rises close to its original value. The maximum temperature in the section is displayed by the red curve. Its sudden rise indicates when the crack has arrived.

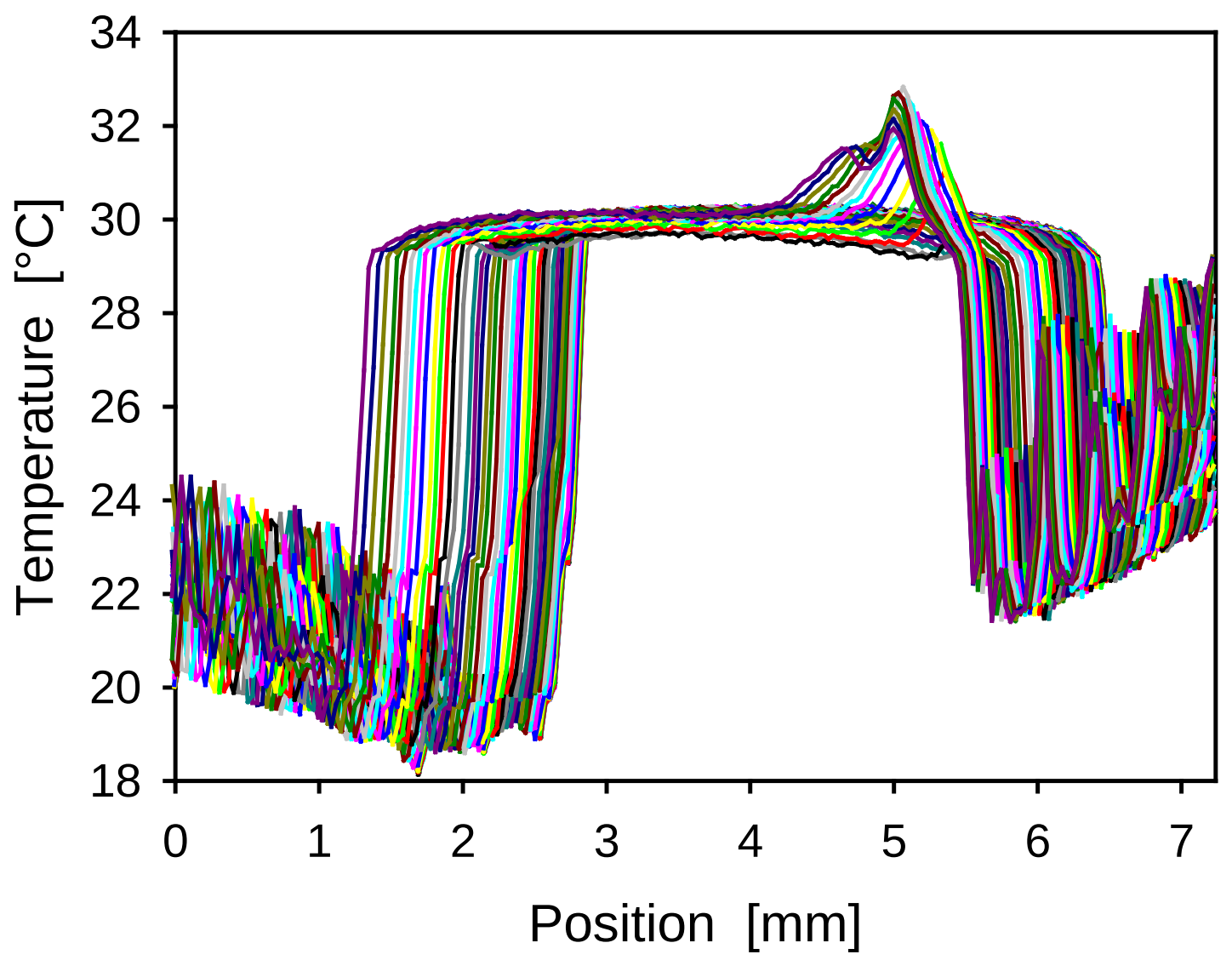

Figure 13: Apparent temperature profile along one cross section for different time points

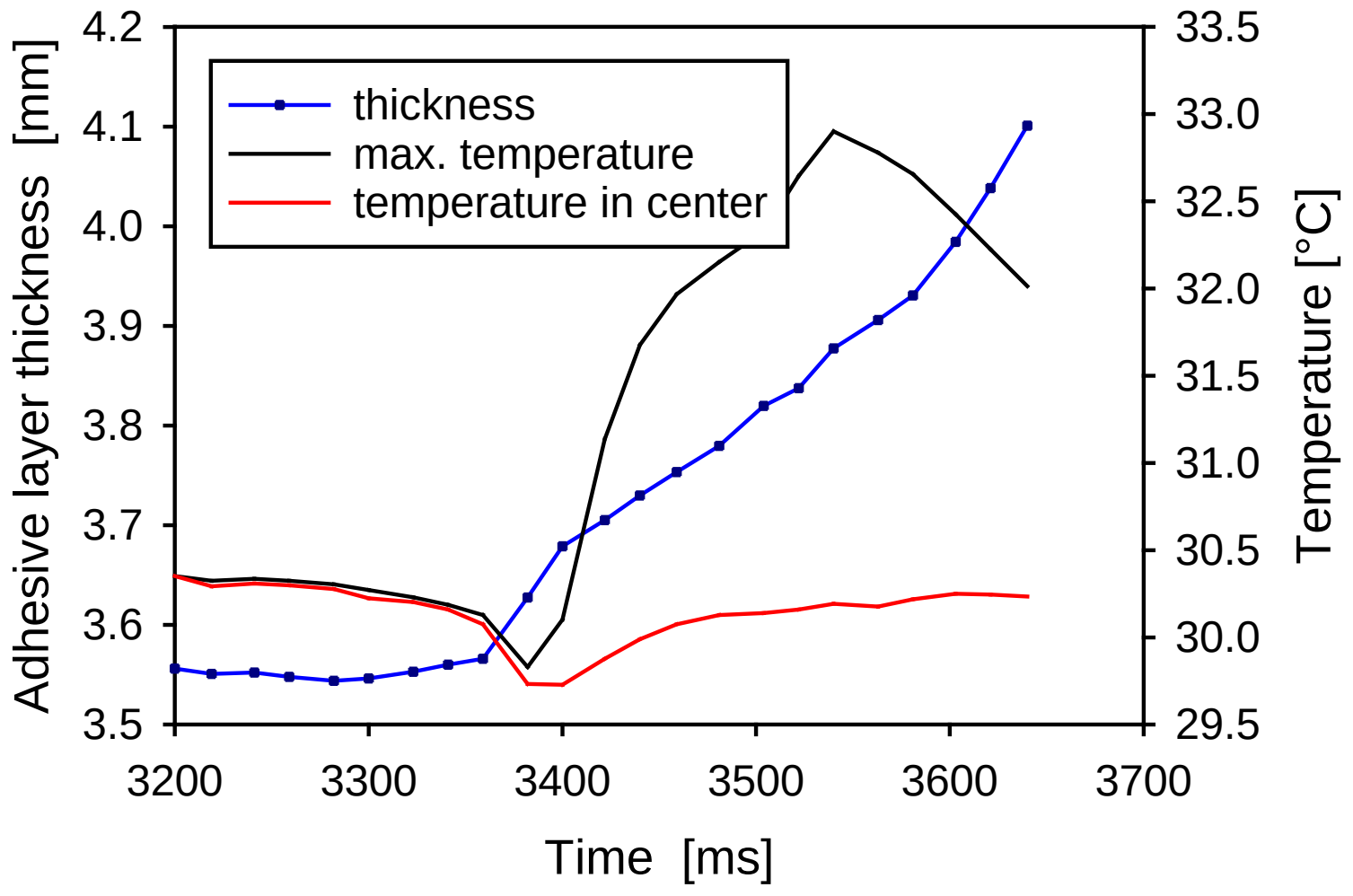

Figure 14: Evolution of local thickness and temperature in time

## 3.2 Thermo-elastic effect

The decrease of temperature during loading of the adhesive layer can be explained by the thermo-elastic effect. Thomson (1878) showed that a (hydrostatic) stress of a solid induces a reversible decrease of temperature [14]:

$$\Delta \boldsymbol{T} = -\frac{\boldsymbol{\alpha}}{\boldsymbol{\rho c_p}} \boldsymbol{T}\, \Delta \mathrm{Tr}\, \boldsymbol{\sigma} = -\boldsymbol{K_0}\, \boldsymbol{T}\, \Delta \mathrm{Tr}\, \boldsymbol{\sigma} \tag{2}$$

Here, $T$ denotes the absolute temperature, $\alpha$ the thermal expansion coefficient, $\rho$ the density, $c_p$ the specific heat, $K_0$ the thermo-elastic constant, and Tr the trace of a tensor.

Considering the temperature dependence of the elastic modulus *E* for plastics, later work explained a quadratic dependence of the temperature change on the stress leading to a stress dependent thermo-elastic constant (e.g. [15],[16])

$$\boldsymbol{K} = \frac{\mathbf{1}}{\boldsymbol{c_\varepsilon \rho}}\left(\boldsymbol{\alpha} - \frac{\mathbf{1}}{\boldsymbol{E}^2}\frac{\boldsymbol{\partial E}}{\boldsymbol{\partial T}}\mathrm{Tr}\,\boldsymbol{\sigma}\right) \tag{3}$$

We performed tensile tests on samples of Betamate 2098 with temperature measurement by infrared camera to identify the thermo-elastic behavior. The temperature measurements were corrected for changes of ambient temperature by direct comparison with an unloaded sample. Furthermore, the effect of heat transfer to the surroundings was corrected based on the cool down curve of an unloaded sample heated in an oven. Figure 15 shows the measured quantities for one of the tensile tests as a function of time. The temperature evaluated at different spots on the sample exhibits significant inhomogeneity. In the range of linear elastic adhesive behavior, the temperature is decreasing. The data can be fitted by a quadratic dependence of the temperature change on the stress. During the plastic deformation of the adhesive, the thermo-elastic effect is overlaid by a temperature rise due to the dissipation in the plastic deformation. For metals models usually assume that a certain fraction (typically 90%) of the plastic work is transformed into heat. If we attempt a similar approach for our adhesive, we obtain a good fit of the temperature change by the model assuming that 18% of the plastic work generates heat. This fraction is much smaller than for metals. A possible explanation is the larger role of visco-elastic effects: a large fraction of the work during deformation is stored in the adhesive and will probably be dissipated during relaxation on a time scale larger than the duration of the performed tensile test.

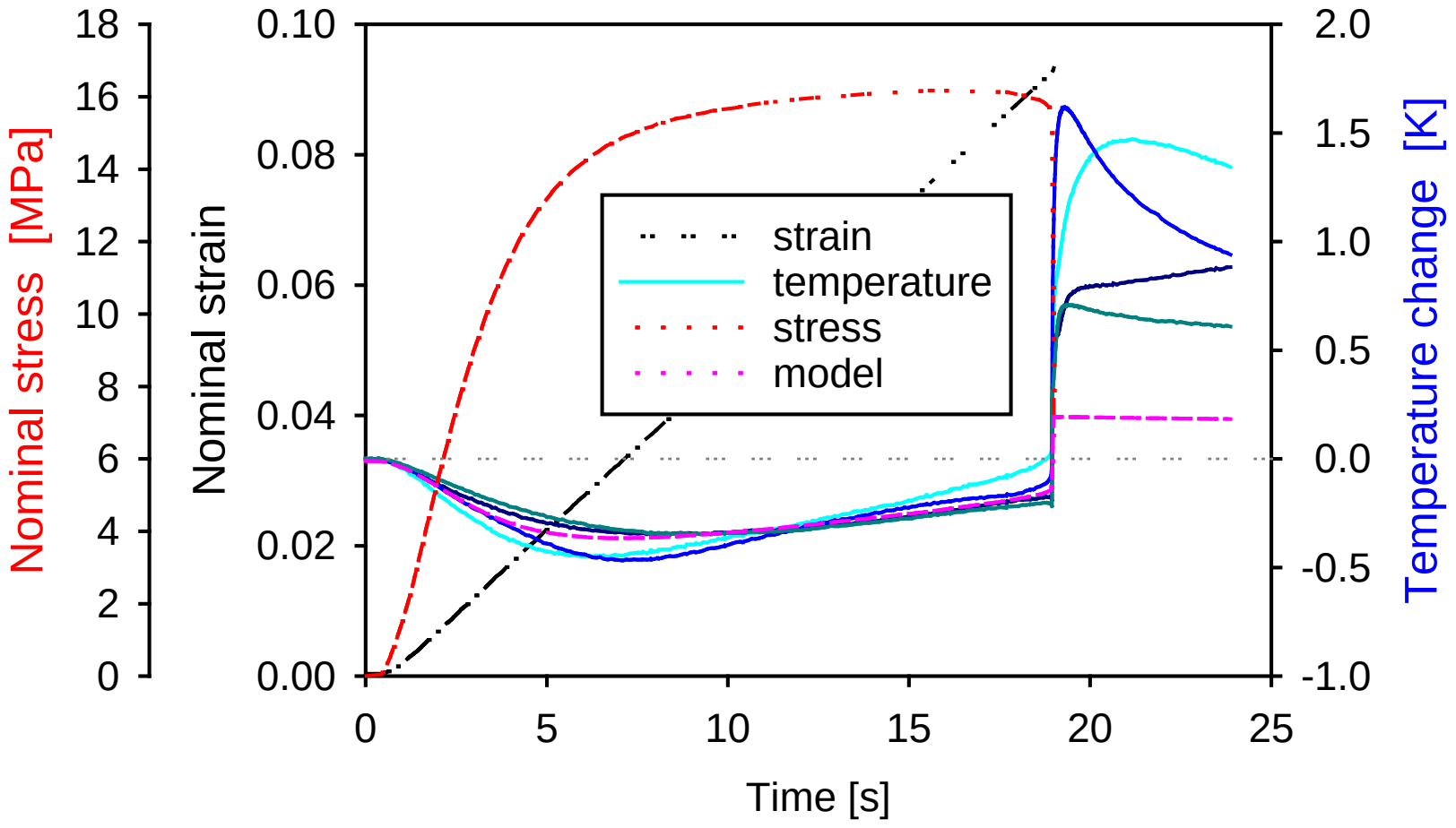

**Figure 15: Result of tensile test with temperature measurements**

A closer inspection of Figure 14 shows that the temperature in the adhesive layer decreases by approximately 0.2 °C while the adhesive is strained by 0.6% in thickness direction. If we assume that at the free surface observed by the infrared camera there is no stress normal to the surface and that the strain in direction of crack propagation is approximately zero, then we can estimate a hydrostatic stress of 3 MPa corresponding to the 0.6% strain. With the thermo-elastic behavior observed in the tensile test, this stress should lead to a temperature decrease of 0.2 K which fits to the measurement results in the TDCB test. Thus, the temperature change in the adhesive layer far from the crack can be explained quantitatively by the thermo-elastic effect.

## 3.3 Estimate of generated heat

This leaves the temperature rise close to the crack to be evaluated quantitatively. A detailed analysis of the infrared video indicates that the high apparent temperatures in the crack region of the last picture in Figure 12 are not all due to a local increase of the adhesive surface temperature. At the stage the crack has opened wide so that the observer is not only looking onto the original adhesive layer surface but also into the opened crack. This means the contrast is not only due to temperature change but also a geometrical contrast. An analysis of pictures and profiles allows estimating the position of the crack edge. The temperature on the surface must be integrated up to that edge to calculate the heat generated at the crack. For example, in Figure 16 the cyan curve displays the temperature distribution at one section immediately before fracture of the specimen. The vertical, dashed cyan line marks the left edge of the crack. The area under the curve over the horizontal, dashed line should be integrated to calculate the heat.

Since the infrared camera is only able to measure the surface temperature and not the temperature in the interior of the adhesive layer, we can obtain an estimate of the heat generated during creation of a unit area of crack only if we assume a homogeneous temperature across the width of the TDCB specimen. Assuming further that we can neglect any dependence of the specific heat on the strain of the adhesive, the heat Q per area of crack growth A is

$$\frac{Q}{A} = c_p \, \rho \int \Delta T \, dy \qquad (4)$$

With the specific heat and density measured for Betamate 2098 ($c_p = 1.51\, Jg^{-1}K^{-1},\ \rho = 1.15\ g\ cm^{-1}$) the heat generated on the left side of the crack is approximately $0.77 \pm 0.16\ kJ/m^2$ where the standard deviation is calculated from the results at different sections. Unfortunately, it is not possible to evaluate the heat on the other side of the crack as well, because it is too close to the adherend. Assuming that a similar amount of heat is generated on both sides of the crack, the total heat per crack area can be estimated to be $1.5\ kJ/m^2$. For comparison, the critical energy release rate of the same TDCB specimen was $3.4\ kJ/m^2$.

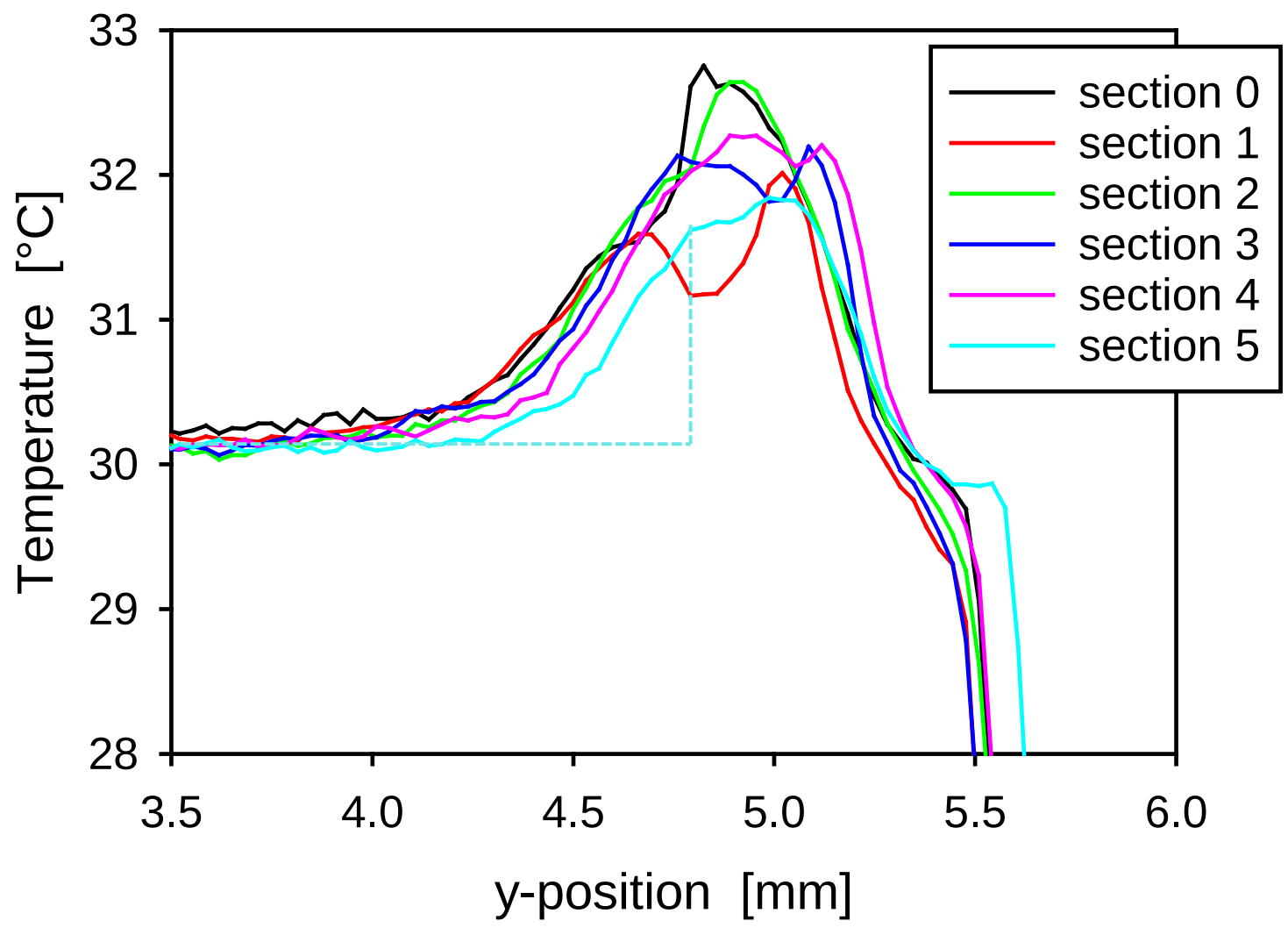

Figure 16: Temperature profile close to crack immediately before total fracture

# 4 Discussion

Based on the results of the simulations and of the temperature measurements a first estimate of the energy balance in the TDCB test can be attempted. The balance is set up for the adhesive Betamate 2098, adhesive thickness 3.6 mm and test velocity 8 mm/s.

One side of the balance is the critical energy release rate as identified by the standard evaluation of the TDCB test, in this case 3.4 $kJ/m^2$. This contains the external work done on the specimen minus the elastic energy stored in the adherends.

On the other side we have several contributions, one of them the kinetic energy. For the considered test velocity this is several orders of magnitude smaller than $G_{Ic}$.

According to the simulations, the plastic work performed in the adhesive layer apart from the close vicinity of the crack is about 82% of $G_{Ic}$. Based on the results of the tensile tests with temperature measurement, this contribution to the balance can be split into stored energy (67% of $G_{Ic}$) and immediately generated heat (15% of $G_{Ic}$).

The observations of the TDCB tests with infrared camera lead to the estimate that about 44% of $G_{Ic}$ appears as heat close to the crack. A more detailed analysis shows that a part of this heat is already accounted for by the estimate for the heat generated by plasticity, so that the total has to be reduced by 8.

In sum we have 118% of $G_{Ic}$, so that the terms not accounted for yet must contribute -18%. One contribution which cannot be estimated with the proposed methods is the energy stored by plastic work performed in the vicinity of the crack, because the exact material behavior at high strains is unknown. A second contribution is the energy required to form the free surfaces of the crack. Both contributions have positive sign. A negative term can arise from the oxidation of the crack surfaces.

This way the methods proposed in this paper can help to estimate an energy balance of the crack growth in the TDCB test. It must be emphasized, however, that this estimate bases on certain approximations and assumptions. These limitations shall be summarized now:

- The simulations require stable crack growth. An adhesive exhibiting stick-slip behavior cannot be treated by this method.
- The model prescribes the crack path parallel to the adherend-adhesive interface. In reality, many adhesives show a less smooth fracture pattern. Although the direction of crack growth, directional instabilities and alternating cracks have been subject of many studies, the incorporation of these effects in the simulations presented here is still very challenging.
- The simulation assumes the existence of a unique crack which is not valid for all adhesives. It was observed that cracks initiate at several places and become connected during further loading.
- The simulations are set up in two dimensions only. Considering that quite little performance was gained by parallel execution on several processors, the generalizations to three dimensions seems not to be feasible today. A three dimensional model might be necessary if the approximation of a straight crack front is far from the reality.
- The infrared camera provides the surface temperature but no direct information about the temperature deep inside the adhesive layer. The evaluation assumed that the temperature is constant across the width of the specimen. Considering that the hydrostatic stress close to the surface differs much from its value in the center of the adhesive layer, it seems not unlikely that the heat created by the crack close to the surface differs from the heat generated inside the adhesive layer.

- If the crack grows close to an adherend, the heat generated on the side of the adherend is difficult to observe. Therefore, it was assumed that the same amount of heat is produced on both sides of the crack. If this is not the case, this simplification can cause an error up to a factor of two.
- The measurements with the infrared camera were applied to specimens of different adhesive thicknesses. So far, only for thick adhesive layers a quantitative evaluation leading to the estimate of heat generation was successful.
- The evaluation of tensile tests with temperature measurement lead to the estimate that 18% of the plastic work is transformed into heat. The question on which time scale this transformation takes place and how this relates to the time scale in the TDCB test has not yet been addressed in detail.
- The specific heat is required to calculate the heat from the measured temperature. This property was determined from a mechanically unloaded specimen, but could change if the adhesive is strained.
- The temperature is calculated assuming that the emission coefficient of the adhesive is constant during the test. Since the observed temperature changes are small, small changes of emission coefficients can cause the same change of measured emission. An apparent temperature change of 1 °C could be caused by a relative change of the emission coefficient by 1.3%. It seems possible that microcracking close to the macroscopic crack influences the emission coefficient.

# 5 Conclusions

The critical strain energy release rate measured in the TDCB test contains the entire work required for the crack growth as well as for the inelastic deformation of the adhesive layer. Two methods have been proposed to gain more detailed information how this work is spent. The first approach uses finite element simulations of the TDCB test based on a material model identified from tensile tests. It allows separating the work required for plastic deformation of the major part of the adhesive layer from the work required for the crack growth and the deformation of the adhesive in its immediate vicinity. The second method adds the measurement of adhesive surface temperature by an infrared camera to the TDCB test. It shows the thermo-elastic effect in the adhesive layer as well as heat generation close to the crack. The results of the two investigations have been combined to create a first estimate of an energy balance of the TDCB test. The application to a ductile epoxy adhesive shows the feasibility of the proposed methods.

# 6 References

[1] Chaves FJP, da Silva LFM, de Moura MFSF, Dillard DA, Esteves VHC. Fracture mechanics tests in adhesively bonded joints: a literature review. J Adhes 2014;90:955-92.
[2] ISO 25217:2009. Adhesives - Determination of the mode 1 adhesive fracture energy of structural adhesive joints using double cantilever beam and tapered double cantilever beam specimens.
[3] ASTM D3433 – 99. Standard test method for fracture strength in cleavage of adhesives in bonded metal joints.
[4] Pearson RA, Blackman BRK, Campilho RDSG, de Moura MFSF, Dourado NMM, Adams RD, Dillard DA, Pang JHL, Davies P, Ameli A, Azari S, Papini M, Spelt JK, Nicoli E, Singh HK, Frazier CE, Giannis S, Armstrong KB, Murphy N, Kawashita LF. Quasi-static fracture tests. In: da Silva LFM, Dillard A, Blackman B, Adams RD, editors. Testing Adhesive Joints, Best Practices. Wiley-VCH; Weinheim, 2012.

[5] Tvergaard V, Hutchinson JW. The influence of plasticity on mixed mode interface toughness. J Mech Phys Solids 1993;41:1119-35.
[6] Suo Z, Shih CF, Varias AG. A theory for cleavage cracking in the presence of plastic flow. Acta metal mater 1993;41:1551-7.
[7] Wei Y, Hutchinson JW. Models of interface separation accompanied by plastic dissipation at multiple scales. Int J Fract 1999;95:1-17.
[8] Chowdhury SR, Narasimhan R. A cohesive finite element formulation for modelling fracture and delamination in solids. Sadhana 2000;25:561-87.
[9] Madhusudhana KS, Narasimhan R. Experimental and numerical investigations of mixed mode crack growth resistance of a ductile adhesive joint. Eng Fract Mech 2002;69:865-83.
[10] Pardoen T, Ferracin T, Landis CM, Delannay F. Constraint effects in adhesive joint fracture. J Mech Phys Solids 2005;53:1951-83.
[11] Martiny Ph, Lani F, Kinloch AJ, Pardoen T. A multiscale parametric study of mode I fracture in metal-to-metal low-toughness adhesive joints. Int J Fract 2012;173:105-33.
[12] Martiny Ph, Lani F, Kinloch AJ, Pardoen T. A maximum stress at a distance criterion for the prediction of crack propagation in adhesively-bonded joints. Eng Fract Mech 2013;97:105–35.
[13] Jokinen J, Wallin M, Saarela O. Applicability of VCCT in mode I loading of yielding adhesively bonded joints – a case study. Int J Adhes Adhes 2015;62:85-91.
[14] Thomson W. On the thermoelastic, thermomagnetic and pyro-electric properties of matters. Phil Mag 1878;5:4-27.
[15] Wong AK, Jones R, Sparrow JG. Thermoelastic constant or thermoelastic parameter? J Phys Chem Solids 1987;48:749-53.
[16] Pitarresi G, Patterson EA. A review of the general theory of thermoelastic stress analysis. J Strain Anal Eng Des 2003;38:405-17.